\documentclass[letter,twocolumn,showpacs,preprintnumbers,pra,byrevtex]{revtex4}

\usepackage{graphics}
\usepackage{amsmath}
\usepackage{amssymb}
\usepackage{bm}
\usepackage{graphicx}

\begin{document}
\title{Hierarchy of inequalities for quantitative duality}
\author{Jes\'us Mart\'{i}nez-Linares }\thanks{Electronic address: jesusml@us.es}
\affiliation{{\it Departamento de F\'{i}sica Aplicada II.\\
Universidad de Sevilla. 41012-Seville, Spain.}}

\date{Received June 16, 2006; revised manuscript received 31 January 2007}

\begin{abstract}
We derive
different relations quantifying duality in a generic two-way
interferometer. These relations set different upper bounds to the
visibility $\mathcal{V}$ of the fringes measured at the output port
of the interferometer. A hierarchy of inequalities is presented
which exhibits the influence of the availability to the experimenter
of different sources of which-way information contributing to the
total distinguishability $\mathcal{D}$ of the ways.
For mixed states and unbalanced interferometers an inequality is
derived, $\mathcal{V}^2+ \Xi^2 \leq 1$, which can be more stringent
than the one associated with the distinguishability ($\mathcal{V}^2+
\mathcal{D}^2 \leq 1$).

%
%
\end{abstract}



\maketitle

%
%
%

\section{INTRODUCTION}
The principle of complementarity is one of the mayor cornerstones of
Quantum Mechanics \cite{Feynman65}. The implications of this
principle
continue to nurture debates in quantum theory \cite{Einstein}. A new
inequality has been introduced \cite{todos,Jaeger95,Englert96} that
quantifies the notion of duality in the context of two-way
interferometers. This inequality has attracted great interest both
theoretically \cite{Bjork98, Englert2000} and experimentally
\cite{DurrNature98, Englert99, Durr2000}.
The inequality establishes an
upper bound to the fringe visibility $\mathcal{V}$ displayed by a
two-level system (the ``quanton") at the output port of a two-way
interferometer. This bound is given by the
\textit{distinguishability} $\mathcal{D}$, i.e., the maximum amount
of which-way information (WWI) that can be potentially available to
the experimenter \cite{Englert96}; namely,
\begin{equation}
\mathcal{V}^2 \le 1-\mathcal{D}^2 .
\label{O1}
\end{equation}

However, two different sources of WWI contributes to $\mathcal{D}$.
One is the a-priori WWI given by the \textit{predictability}
$\mathcal{P}$ of the ways, i.e., the \textit{a priori} which-way
knowledge that the experimenter has about the ways stemming from the
preparation of the beam splitter (BS) and the initial state of the
quanton. $\mathcal{P}$  is a measure of WWI on its own. In fact, it
satisfies the inequality \cite{todos,Englert96}
\begin{equation}
\mathcal{V}^2 \le 1-\mathcal{P}^2 . \label{O2P}
\end{equation}
In addition, the experimenter may place a quantum memory system to
interact with the quanton in order to acquire extra WWI, i.e., to
serve as a which-way marker (WWM). Thus, the second source of WWI
stems from the WWM's ability to correlate its final states with the
two ways, leading to the storage of some WWI. A measure of the
quantum ``quality" $\mathcal{Q}$  of the WWM has been introduced in
\cite{Jesus04}. We will show that this quantity also obeys an
inequality of the same type, i.e.,
\begin{equation}
\mathcal{V}^2 \le  1-\mathcal{Q}^2 . \label{O2Q}
\end{equation}

Moreover, the final visibility is limited by the availability of
both kinds of WWI by the inequality
\begin{equation}
\mathcal{V}^2 \le \left(1-\mathcal{P}^2\right)
\left(1-\mathcal{Q}^2\right) . \label{O2}
\end{equation}

\begin{figure}
\includegraphics[scale=1.1]{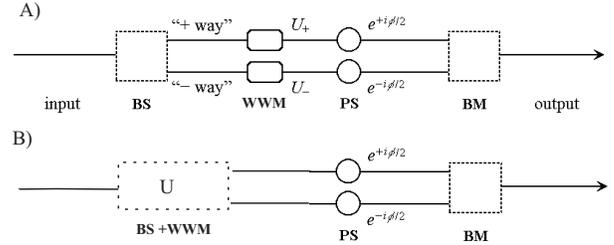}
\caption{\label{fig1}  Schematic two-way interferometer setup
composed of a beam splitter (BS), a quantum which-way marker (WWM),
a phase shifter (PS), and a beam merger (BM). (a) The WWM is
characterized by unitary $U_{\pm}$ evolution. (b) Quantized BS: in
this general case the same physical interaction is used for both
splitting the beam and marking the way.}
\end{figure}


Equation (\ref{O2}) has been derived in \cite{Jesus04} for the
special case of unitary WWM evolution [Fig.1(a)]. Here we will prove
it for a generic two-way interferometer, i.e., the general case
where WWM evolution can be non unitary, so quantum optical Ramsey
interferometers \cite{newHaroche2001} can also be analyzed
[Fig.1(b)]. Then, we show that for two-level WWM systems Eq.
(\ref{O2}) can be more stringent than Eq. (\ref{O1}).

This scheme sets the organization of the paper. First we describe in
Sec. \ref{section2} the formalism for generic two-way
interferometers with non-unitary WWM. Section \ref{section3} is
devoted to the derivation of Eqs. (\ref{O2Q}) and (\ref{O2}). In
Sec. \ref{section4} Eq. (\ref{O2}) is interpreted as an inequality
quantifying duality. Here we also prove the stringency statement. In
Sec. \ref{sectionSQDS} the results are illustrated with the help of
a simple example: the symmetric quantum-detecton system (a
particular quantum logic gate). Finally we end up with conclusion
and a summary of the results.

\section{NONUNITARY WWM}
\label{section2}
 Let us consider the generic two-way interferometer
plotted in Fig.1(b). Following the notation of \cite{Jesus04} the
quanton is prepared initially in the state
\begin{equation}
\rho_Q^{(0)} = \frac{1}{2} \left( 1 + \boldsymbol{s}_Q^{(0)} \cdot
\boldsymbol{\sigma} \right),  \label{2}
\end{equation}
where $\boldsymbol{\sigma} = (\sigma_x,\sigma_y,\sigma_z) $ are the
usual Pauli spin operators, and $\boldsymbol{s}_Q^{(0)} = (0, 0, s)
$ is the polarization vector of the state, characterized by the
inversion $s\in (-1,1)$.  We now consider the most general case and
include BS and WWM action into the same global action characterized
by the operator $U$.
The combined quanton-WWM system is initially prepared in the state
$\rho^{(0)} = \rho_Q^{(0)} \otimes \rho_D^{(0)}$, where
$\rho_D^{(0)}$ is an arbitrary initial state of the WWM. After
interaction between the quanton and WWM, the evolution of the entire
system is given by the unitary map $\rho \rightarrow U^{\dag } \,
\rho  U$.
We follow Englert's notation \cite{Englert96b} and write the general
evolution operator in the quanton $\sigma _{z}$ base as
\begin{equation}
\label{O6}
 U = \frac{1}{\sqrt{2}} \left(
\begin{array}{cc}
V_{++} & V_{+-} \\
-V_{-+} & V_{--}
\end{array}
 \right),
\end{equation}
with the matrix elements acting exclusively on the WWM. These four
operators are not necessarily unitary, though they are restricted by
the unitarity of $U$. Note that the particular case  $V_{++} =
V_{-+}=U_+$, $V_{+-} = V_{--}=U_-$, with
$U_{\pm}U_{\pm}^\dagger=U_{\pm}^\dagger U_{\pm}=I$ brings us back to
the unitary case of Fig.1(a) studied in \cite{Jesus04}. Thus, the
results of this paper cover both situations schematized in Figs.
1a-b.


Now is the turn of the phase shifter (PS), that effects the
transition
\begin{equation}
\rho _{Q}\rightarrow \exp \left( -i\frac{\phi }{2}\sigma _{z}\right)
 \rho_{Q} \; \exp \left( i\frac{\phi }{2}\sigma _{z}\right) ,
\label{O7}
\end{equation}
and subsequently the beam merger (BM)
\begin{equation}
\rho _{Q}\rightarrow \exp \left( -i\frac{\pi }{4}\sigma _{y}\right)
 \rho_{Q} \; \exp \left( i\frac{\pi }{4}\sigma _{y}\right) .
\label{4}
\end{equation}
Combining transformations (\ref{O6}), (\ref{O7}) and (\ref{4}), the
general final state \cite{comentario inversion} at the output port
of the  interferometer given in Fig.\ref{fig1}(b) is
\begin{equation}
\label{O8} \rho^{(f)} =\frac{1+s}{2} \; \rho^{(f)}_\uparrow
+\frac{1-s}{2}\; \rho^{(f)}_\downarrow ,
\end{equation}
where
\begin{eqnarray}
\rho^{(f)}_\uparrow &=&\mbox{$\frac{(1+\sigma_{x})}{4}$} \;
               V_{++}^{\dag }\rho _{D}^{(0)}V_{++}
            + \mbox{$\frac{(1-\sigma _{x})}{4}$}  \;
                V_{+-}^{\dag }\rho _{D}^{(0)}V_{+-}
\nonumber \\
            &+&\mbox{$\frac{ (-\sigma_{z}+i\sigma_{y})}{4}$}
           \, V_{++}^{\dag }\rho _{D}^{(0)}V_{+-}\;e^{-i\phi }
\nonumber \\
             &-&\mbox{$\frac{(\sigma_{z}+i\sigma_{y})}{4}$}
              \, V_{+-}^{\dag }\rho_{D}^{(0)}V_{++}\;e^{i\phi }\, ,
\label{7}
\end{eqnarray}
and $\rho^{(f)}_\downarrow$ can be obtained from the above equation
through the replacements
\begin{equation}
V_{++}\rightarrow -V_{-+}; \:\; V_{+-}\rightarrow V_{--}\;.
\label{replacements}
\end{equation}
After tracing over the WWM's degrees of freedom, the final state of
the quanton can be expressed in terms of a Bloch vector
$\boldsymbol{\sigma}^{(f)}$ with components
\begin{eqnarray}
S_{Qx}^{(f)} &=& w_+ -w_- \, ,
\nonumber \\
S_{Qz}^{(f)} &=&  -\frac{1+s}{2} \; \Re e \left[\mathcal{C}_\uparrow
e^{-i\phi}\right] - \frac{1-s}{2} \; \Re e
\left[\mathcal{C}_\downarrow e^{-i\phi}\right],
\nonumber\\
S_{Qz}^{(f)}+i S_{Qy}^{(f)} &=& - \;e^{-i\phi }\mathcal{C} ,
\label{O10}
\end{eqnarray}
where
\begin{equation}
w_{\pm} =\mbox{tr} \left\{\frac{1\pm \sigma_x}{2}
\rho^{(f)}\right\},
 \label{O11}
\end{equation}
$w_++w_-=1$, are the probabilities for taking the two alternative
ways at the central stage of the interferometer,
and
\begin{equation}
 \mathcal{C} = \frac{1+s}{2} \; \mathcal{C}_{\uparrow} + \frac{1-s}{2}\;
 \mathcal{C}_{\downarrow} ,
 \label{O12}
\end{equation}
with
\begin{eqnarray}
 \mathcal{C}_{\uparrow}      &\equiv& \mbox{tr}_D
              \, \left\{ V_{++}^{\dag }\rho_{D}^{(0)}V_{+-}\right\}= \left< V_{+-} V_{++}^{\dag }
              \right>_0 ,
\nonumber \\
 \mathcal{C}_{\downarrow}     &\equiv&  -\mbox{tr}_D
              \, \left\{ V_{-+}^{\dag }\rho_{D}^{(0)}V_{--}\right\}= -\left< V_{--} V_{-+}^{\dag } \right>_0 .
 \label{O13}
\end{eqnarray}

Inserting Eq. (\ref{O8}) into Eq. (\ref{O11}) the transition
probabilities are calculated to be
\begin{eqnarray}
w_+ &=&\frac{1+s}{4} \;
                \left< V_{++} V_{++}^{\dag } \right>_0 +\frac{1-s}{4} \;
                \left< V_{-+} V_{-+}^{\dag } \right>_0 ,
\nonumber\\
w_- &=&\frac{1+s}{4} \;
                \left< V_{+-} V_{+-}^{\dag } \right>_0 +\frac{1-s}{4} \;
                \left< V_{--} V_{--}^{\dag } \right>_0 .
 \label{O14}
\end{eqnarray}
The mean value of their difference is the predictability of the
ways, i.e.,
\begin{equation}
\mathcal{P}= |w_+ -w_-|. \label{OP}
\end{equation}
$\mathcal{P}$ is a-priori WWI.
The x component of the Bloch vector given in (\ref{O10}) represents
{\it a priori} WWI.
The other components give information about complementary wavelike
aspects of the quanton. In order to see this, let us focus on the
output port of the interferometer. The probability of measuring the
upper or lower state of the quanton, after many repetitions of the
experiments, displays an interference pattern versus variation of
the phase in the PS.
The visibility of the fringes can be rapidly calculated with the
help of (\ref{O10}) to yield
\begin{equation}
\mathcal{V} = \mid \mathcal{C}\mid\le 1 . \label{O16}
\end{equation}
$\mathcal{C}$ is therefore a contrast factor.
Inserting Eqs. (\ref{O16}) and (\ref{OP}) into Eq. (\ref{O10}), it
is easy to check that the duality relation given in Eq. (\ref{O2P})
is also valid for non-unitary WWM.

More stringent inequalities than (\ref{O2P}) can be derived for the
system. Englert \cite{Englert96b} derives the inequality given in
Eq. (\ref{O1}) for the distinguishability
\begin{equation}
\mathcal{D}=\text{tr}_{D}\{|w_+ \rho _{D}^{(+)}-w_- \;\rho
_{D}^{(-)}|\}, \label{OD}
\end{equation}
where
\begin{eqnarray}
w_+ \rho_D^{(+)}&=& \text{tr}_Q \left\{\frac{1+\sigma_x}{2} \;
\rho^{(f)}\right\}
\nonumber\\
&=& \frac{1+s}{4}\;
                 V_{++}^{\dag} \rho_D^o V_{++}  +\frac{1-s}{4} \;
                 V_{-+}^{\dag} \rho_D^o V_{-+}   ,
\nonumber\\
w_- \rho_D^{(-)}&=& \text{tr}_Q \left\{ \frac{1-\sigma_x}{2} \;
\rho^{(f)} \right\}
\nonumber\\
&=& \frac{1+s}{4} \;
                 V_{+-}^{\dag} \rho_D^o V_{+-}  +\frac{1-s}{4} \;
                 V_{--}^{\dag} \rho_D^o V_{--}  .
 \label{O20}
\end{eqnarray}
$\rho_D^{(\pm)}$ are the final states of the WWM associated with a
measure of the $\sigma_z=\pm 1$ way \cite{comentario BM}. These
states are normalized, so $\rho_D^{(f)}= \text{tr}_Q \; \rho^{(f)}=
w_+ \rho_D^{(+)}+ w_- \rho_D^{(-)}$  is the final state of the WWM
provided no measure of the ways is taken.

Two kinds of WWI are represented in $\mathcal{D}$. One is the
predictability of the ways, given in (\ref{OP}). The other one is
given by the quantum ``quality" of the WWM, i.e., its
ability to establish quantum correlations with the quanton, leading
to the storage of some WWI into the final state of the WWM. In
\cite{Jesus04} the quality measure
\begin{equation}
\mathcal{Q}= \frac{1}{2}\;\text{tr}_{D}\{\mid \rho _{D}^{(+)}-\rho
_{D}^{(-)}| \} \label{OQ}
\end{equation}
 has been introduced. $\mathcal{Q}$ is a distance between the conditional probabilities
 $\rho _{D}^{(+)}$ and $\rho _{D}^{(-)}$ in the trace-class norm.
Note that for balanced interferometers ($\mathcal{P}=0$),
 $\mathcal{Q}$ coincides with $\mathcal{D}$. Otherwise their value
 may substantially differ. In both cases  $\mathcal{Q}$ is a quantitative measure of the WWM's intrinsic ability to distinguish between
the quanton's alternatives. For $\mathcal{Q}=0$  the marker cannot
distinguish the ways at all. Conversely, full WWI can be stored in
the marker when $\mathcal{Q}=1$.
The states $\rho _{D}^{(\pm)}$ can be prepared if we modify the
interferometer such that the actual way taken by the quanton is
measured rather than the fringe pattern.
Thus, the value of $\mathcal{Q}$ can be experimentally measured
along the same lines described in \cite{Englert96,Englert99} for
measuring $\mathcal{D}$.

\section{More inequalities}
\label{section3}
 In \cite{Jesus04} we restricted the analysis for the particular case of unitary matrix elements in
 (\ref{O6}), so (\ref{OQ}) is a distance between projectors.
 Here, we release this condition and prove the validity of Eq. (\ref{O2}) in
the general case.

The validity of Eq. (\ref{O1}) for nonunitary WWM has been given by
Englert in \cite{Englert96b}. Following Englert, we separate first
both contributions on the right side of Eqs. (\ref{O20}) so
\begin{equation}
\mathcal{Q}_{\uparrow}= \frac{1}{4}\;\text{tr}_{D} \left\{  \left|
\frac{V_{++}^{\dag}\rho
_{D}^{(0)}V_{++}}{w_+}-\frac{V_{+-}^{\dag}\rho
_{D}^{(0)}V_{+-}}{w_-} \right| \right\},
 \label{O26}
\end{equation}
gives the value of the quality $\mathcal{Q}$ for the $s=1$ case. For
the $s=-1$ case, $\mathcal{Q}_{\downarrow}$ can be obtained from
(\ref{O26}) by taking the replacements (\ref{replacements}). The
triangle inequality $\text{tr}\left\{ \mid X+Y\mid \right\}\le
\text{tr}\left\{ \mid X\mid \right\}+\text{tr}\left\{ \mid Y\mid
\right\}$ applied to (\ref{OQ}) yields
\begin{equation}
\mathcal{Q}\le \frac{1+s}{2}\; \mathcal{Q}_{\uparrow} +
\frac{1-s}{2}\; \mathcal{Q}_{\downarrow}. \label{O27}
\end{equation}

According to this notation and using (\ref{O16}), the duality
relation given in (\ref{O2}) can be written for the particular cases
$s=1$ ($\uparrow$) and $s=-1$ ($\downarrow$) as
\begin{equation}
\mathcal{Q}_{\uparrow\downarrow}^2
+\frac{|C_{\uparrow\downarrow}|^2}{1-\mathcal{P}^2}\le 1
.\label{O28}
\end{equation}
Here, we prove the $s=1$ case. The $s=-1$ case will follow
straightforwardly, once the replacements (\ref{replacements}) are
taken. In order to do this, let us define the traceless operator
\begin{equation}
\Gamma = \frac{1}{2}\; \{\rho _{D}^{(+)}-\rho _{D}^{(-)}\},
\label{O30}
\end{equation}
Its eigenvalues are of the form $\pm\lambda$, so we can write
\begin{equation}
\mathcal{Q} =2|\lambda| . \label{O31}
\end{equation}
To calculate $\lambda$, we use the relation $\text{tr}_D\,\Gamma^2
=2\lambda^2$, and
\begin{eqnarray}
4\text{tr}_D\,\Gamma^2 &=&\text{tr}_D  \{\rho_D^{(+)2}\}
+\text{tr}_D
\{\rho_D^{(-)2}\} \nonumber\\
&-&\text{tr}_D \{\rho_D^{(+)}\rho_D^{(-)}\}-\text{tr}_D
\{\rho_D^{(+)}\rho_D^{(-)}\}.
 \label{O31bis}
\end{eqnarray}
We particularize for the WWM's pure state preparation, where we can
work out explicitly the above four contributions. In fact, each of
the first two terms is unity, in both the unitary and nonunitary
cases.
Next, we calculate the cross terms in (\ref{O31bis}). From the
definitions (\ref{O20}) and (\ref{O13}) we have
\begin{equation}
\text{tr}_D \{\rho_D^{(+)} \rho_D^{(-)}\}= \frac{\mid
C_{\uparrow}\mid^2}{4w_+ w_- }=\text{tr}_D \{\rho_D^{(-)}
\rho_D^{(+)}\}.
 \label{O33}
\end{equation}
Collecting terms, we obtain
\begin{equation}
2 \,\text{tr}_D \,\Gamma^2 =1-\frac{\mid
C_{\uparrow}\mid^2}{1-\mathcal{P}^2} ,
 \label{O34}
\end{equation}
where the relation $1-\mathcal{P}^2=4w_+w_-$ has been used. Finally,
using (\ref{O34}), (\ref{O31}) and (\ref{replacements}) we obtain
the pure state result
\begin{equation}
\mathcal{Q}_{\uparrow\downarrow}^2 +\frac{\mid
C_{\uparrow\downarrow}\mid^2}{1-\mathcal{P}^2} =1 .
 \label{O35}
\end{equation}
Now consider the general case where we prepare the WWM in a mixed
state. Its spectral decomposition allow us to write it as a
combination of pure states for which (\ref{O35}) can be applied;
namely,
\begin{equation}
\rho_D^{(0)} =\sum_{k=1}^N D_k \mid d_k \rangle \langle d_k\mid ,
 \label{O36}
\end{equation}
where $\sum_k D_k =1$ and $\langle d_k \mid d_j
\rangle=\delta_{kj}$. Inserting (\ref{O36}) into (\ref{O26}), and
applying the triangular inequality we obtain
\begin{equation}
\mathcal{Q}_{\uparrow} \le \sum_k D_k \;\mathcal{Q}_{\uparrow k}  ,
 \label{O38}
\end{equation}
where
\begin{eqnarray}
\mathcal{Q}_{\uparrow k}&=& \text{tr}_{D} \left\{ \left|
\frac{V_{++}^{\dag}  \mid d_k \rangle \langle d_k\mid
V_{++}}{w_+}-\frac{V_{+-}^{\dag}\mid d_k \rangle \langle d_k\mid
 V_{+-}}{w_-} \right| \right\},
 \nonumber\\
 \label{O39}
\end{eqnarray}
is the quality for pure state preparation and, thus, obeys
(\ref{O35}). Therefore, Eq. (\ref{O38}) can be written as
\begin{equation}
\mathcal{Q}_{\uparrow} \le \sum_k D_k \;\sqrt{1-\frac{\mid
C_{\uparrow k}\mid^2}{1-\mathcal{P}^2} }\; ,
 \label{O40}
\end{equation}
where $C_{\uparrow k}=\langle d_k \mid V_{+-}V_{++}^{\dag}\mid d_k
\rangle$. We can also calculate $C_{\uparrow}$ in terms of pure
state contributions. In fact, inserting (\ref{O36}) into (\ref{O13})
one obtains $C_{\uparrow} = \sum_k D_k \;C_{\uparrow k}$. Combining
the two last summations, one gets
\begin{eqnarray}
&&\mathcal{Q}_{\uparrow}^2+\frac{\mid
C_{\uparrow}\mid^2}{1-\mathcal{P}^2} \le \sum_{kj} D_k D_j \;
\nonumber\\
&&\times \left[\sqrt{1-\frac{\mid C_{\uparrow
k}\mid^2}{1-\mathcal{P}^2}} \; \sqrt{1-\frac{\mid C_{\uparrow
j}\mid^2}{1-\mathcal{P}^2}} + \frac{\mid C_{\uparrow k}\mid  \mid
C_{\uparrow j}\mid}{1-\mathcal{P}^2} \right].
 \label{O42}
\end{eqnarray}
Let us have a closer look on the matrix term in square brackets
$[\;]_{kj}$ in the above equation. Its diagonal terms satisfy
$[\;]_{kk}=1$, as can be checked by inspection.  In order to prove
that the non-diagonal terms are smaller than unity, let us define
the numbers $\theta_k^2=|C_{\uparrow k}|^2/(1-\mathcal{P}^2)$.
Equation (\ref{O35}) implies $0\le \theta_k\le 1$, which are in turn
duality relations of the type of Eq. (\ref{O2P}). In terms of
$\theta_k$, we can write $
[\;]_{kj}=\sqrt{1-\theta_k^2}\sqrt{1-\theta_j^2}+\theta_k\theta_j, $
which does not exceed unity.
The calculation of the upper bound of (\ref{O42}) is now
straightforward, namely,
\begin{equation}
\mathcal{Q}_{\uparrow}^2+\frac{\mid
C_{\uparrow}\mid^2}{1-\mathcal{P}^2} \le \sum_{kj} D_k D_j =1,
\label{O45}
\end{equation}
which completes the proof of the $s=1$ case of Eq. (\ref{O28}). Eq.
(\ref{O2}) follows now directly from Eqs. (\ref{O27}) and
(\ref{O12}). Moreover, Eq. (\ref{O2Q}) follows from Eq. (\ref{O2}),
since $1-\mathcal{P}^2\le 1$.

\section{THE $\mathcal{V}_{\Xi}$ UPPER BOUND}
\label{section4}
We define the symmetric quantity
\begin{equation}
\Xi = \sqrt{\mathcal{Q}^2
+\mathcal{P}^2-\mathcal{Q}^2\mathcal{P}^2}.
 \label{O47}
\end{equation}
In terms of $\Xi$, Eq. (\ref{O2}) can be written as
\begin{equation}
\mathcal{V}^2 +\Xi^2 \le 1  . \label{nuevita}
\end{equation}

Notice the high degree of symmetry inherent in the structure of Eq.
(\ref{O47}). First, $\Xi$ is bounded by $\mathcal{Q}$ and
$\mathcal{P}$. Actually, as can be rapidly checked in Eq.
(\ref{O47}), $\Xi$ satisfies $0\le\Xi\le 1$, $\Xi\ge \mathcal{Q}$,
$\Xi\ge \mathcal{P}$, $\Xi\ge \mathcal{Q}\;\mathcal{P}$, so is equal
or greater than any of its two sources of WWI. This can be
understood since $\Xi$ yields one kind of WWI, $\mathcal{Q}$ (or
$\mathcal{P}$), when the other kind, $\mathcal{P}$ (or
$\mathcal{Q}$), vanishes. For instance, for $\mathcal{P}=0$ we have
$\mathcal{D}=\mathcal{Q}=\Xi$, i.e, for symmetric interferometers
the inequalities given in Eqs. (\ref{nuevita}) and (\ref{O2Q})
reduces to the inequality given in Eq. (\ref{O1}).

Second, $\Xi$ reaches unity when any of its arguments $\mathcal{Q}$
(or $\mathcal{P})$ does,  independently of the value of the other
argument $\mathcal{P}$ (or $\mathcal{Q})$. In order to understand
the importance of this property we note that Eq. (\ref{O47}) equals
Eq. (\ref{OD}) in the case of pure state preparation ($|s|=1$,
$\text{tr}_D \;\rho_D^{(0)2}=1$), i.e.,
\begin{equation}
\mathcal{D}^2=\Xi^2=\mathcal{P}^2+\mathcal{Q}^2(1-\mathcal{P}^2).
\label{newpure}
\end{equation}
This can be shown by noting that in the pure-state case Eq.
(\ref{O35}) turns into
\begin{equation}
\Xi^2+\mathcal{V}^2=1
\label{nuevita2}
\end{equation}
and, on the other hand, Eq. (\ref{O1}) is satisfied as an equality
\cite{Englert96b}. Thus, for pure state preparation $\Xi$ coincides
with $\mathcal{D}$, i.e., the total distinguishability of the ways.
Equation (\ref{newpure}) renders $\mathcal{P}$ and $\mathcal{Q}$ as
two different contributions to the distinguishability. The condition
$\mathcal{P}=1$  (or $\mathcal{Q}=1$) exhausts the amount of WWI
needed to specify the actual path taken by the quanton, so
$\mathcal{D}=1$ in each case, with independence of the value of the
additional source of WWI.

Third, $\Xi$ treats on equal footing both sources of WWI since it is
invariant under the permutation $\mathcal{P}\leftrightarrow
\mathcal{Q}$. This fact, together with Eq. (\ref{nuevita2}), tells
us that for pure state preparation both sources of WWI stand on
equal footing concerning wave-particle duality (WPD) degradation of
fringe visibility.

Finally, Eq. (\ref{nuevita}) yields the following conditions
\begin{eqnarray}
\mathcal{V}=1 & \quad\Rightarrow\quad &
\Xi=\mathcal{P}=\mathcal{Q}=0 ,
\nonumber \\
\mathcal{P}=1 & \quad\Rightarrow\quad & \Xi=1, \mathcal{V}=0,  \nonumber \\
\mathcal{Q}=1 & \quad\Rightarrow\quad & \Xi=1, \mathcal{V}=0,  \nonumber \\
\Xi=1 & \quad\Rightarrow\quad & \mathcal{V}=0 . \label{boncho}
\end{eqnarray}
as demanded by duality. Thus, Eq. (\ref{nuevita}), like Eq.
(\ref{O1}), is an inequality quantifying WPD. It devolves the
extreme situations of Eqs. (\ref{boncho}) showing that perfect
fringe visibility and the acquisition of full path knowledge on any
source of WWI are mutually exclusive. Intermediate situations of
partial fringe visibility and partial WWI on each source are also
contemplated in Eq. (\ref{nuevita}). Both equations (\ref{nuevita})
and (\ref{O1}) give quantitative statements about duality.
Specifically, Eq. (\ref{nuevita}) allows one to trace the loss of
coherence in asymmetric interferometers to the two reservoirs of WWI
represented by $\mathcal{P}$ and  $\mathcal{Q}$.

 Now, the main result of the paper. We prove that
there exists a wide class of systems  where
\begin{equation}
\mathcal{V}^2 \le  1-\Xi^2\le 1-\mathcal{D}^2 ,
\label{main}
\end{equation}
i.e., Eq. (\ref{O2}) (left-hand inequality above) can be more
stringent than Eq. (\ref{O1}).
In order to do so, we restrict ourselves to a generic two-level WWM
[$N=2$ in Eq. (\ref{O36})] and prove the inequality $\chi \equiv
\mathcal{D}^2/\, \Xi^2 \le 1$, which is equivalent to the right hand
side of (\ref{main}). We use the relation $\mathcal{D} = \text{Max}
\left\{ \mathcal{P}, \mathcal{R} \right\}$, with $\mathcal{R}^2  =2
\; \text{tr}_D \{\underline{\Delta}\}^2 -\mathcal{P}^2 $ and
$\underline{\Delta} \equiv \omega_+ \rho_D^+ -\omega_- \rho_D^- $,
which can be verified for $N=2$. For $\mathcal{P}\ge \mathcal{R}$,
$\chi \le 1$ follows trivially. For $\mathcal{P}< \mathcal{R}$ we
compute explicitly both $\mathcal{D}$ and $\mathcal{Q}$ by
diagonalizing Eqs. (\ref{OD}) and (\ref{OQ}), respectively, along
the same lines as used in the derivation of Eq. (\ref{O35}). After a
rather lengthy calculation the result simplifies remarkably to the
expression
\begin{equation}
\chi = 1 -\frac{4 D_1 D_2 \mathcal{P}^2}{\Xi^2} \le 1,
  \label{O49}
\end{equation}
where we have taken for simplicity $|s|=1$ and $w_{\pm 12}=w_{\pm
21}^*=0$, $w_{\pm ij}$ being the matrix elements of the operators
involved in Eq. (\ref{O14}).


Equation (\ref{main}) defines two different upper bounds to the
visibility, namely,
\begin{eqnarray}
\mathcal{V}_D &=& \sqrt{1-\mathcal{D}^2},
\label{defi1}\\
\mathcal{V}_{\Xi} &=& \sqrt{1-\Xi^2}. \label{defi2}
\end{eqnarray}
In terms of these bounds, Eqs. (\ref{O1}) and (\ref{nuevita}) can be
rewritten as
\begin{eqnarray}
\mathcal{V} &\le& \mathcal{V}_D , \label{VVD}\\
\mathcal{V} &\le& \mathcal{V}_{\Xi}, \label{VVXi}
\end{eqnarray}
and Eq. (\ref{main}) as
\begin{equation}
\mathcal{V} \le \mathcal{V}_{\Xi} \le \mathcal{V}_D. \label{3V}
\end{equation}

Note that Eq. (\ref{VVD}) does not saturate for mixed state
preparation, so it is possible to find a stricter inequality. In
fact, this is what has been accomplished in Eq. (\ref{3V}). Yet Eq.
(\ref{VVXi}) does not saturate for mixed-state preparation
\cite{comentario D=1}, so stricter inequalities could eventually
still be found.




\section{AN EXAMPLE: THE SYMMETRIC DETECTON-QUANTON SYSTEM}
\label{sectionSQDS} 
\begin{figure}
\includegraphics[scale=1.2]{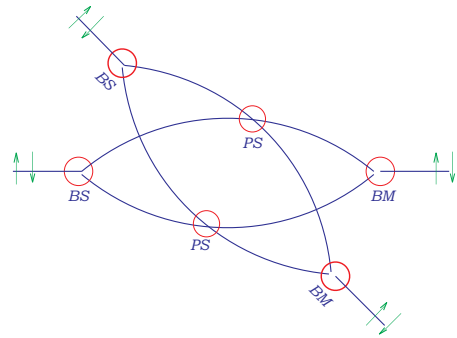}
\caption{\label{figojo}  Schematic setup for the SQDS, showing the
intrinsic symmetry between quanton and detecton. }
\end{figure}

In order to illustrate the formalism, we
particularize it in this section to a concrete example: the
symmetric quanton-detecton system (SQDS) \cite{Jesus04}. The SQDS is
represented in Fig. \ref{figojo}. The system is basically a quantum
logic gate in which each qubit (quanton and detecton) plays the role
of which-way marker of the other. The conditional dynamics is
achieved at the common phase shifter. The detecton phase shifter
depends on the $\pm$ alternative ways of the quanton in the form
\begin{equation}
U_{PS}^{\pm}  = \exp \left( \pm  \frac{i}{2} \; \Phi\;
\sigma_{Dz}\right),
\end{equation}
where $\Phi$ is an entangling phase and $\sigma_{Dz}$ is the usual z
component of the Pauli matrices for the detecton.

 On the other hand, the SQDS can be regarded as a pair of two-way interferometers
coupled at their central stages by a dispersive interaction, trying
to acquire WWI about each other. The detecton, like the quanton
itself, is a two-way interferometer, likewise describable by a
predictability $\mathcal{P}_D$ and a fringe visibility
$\mathcal{V}_{D}^o$ satisfying
\begin{equation}
\mathcal{P}_D^2+\mathcal{V}_{D}^{o2} = |s_D^{(0)}|^2 \le 1,
\label{SDo}
\end{equation}
where $\boldsymbol{s}_D^{(0)}$ is the Bloch vector describing the
initial state of the detecton. We take the quanton prepared
initially in a pure state so
\begin{equation}
\mathcal{P}_Q^2+\mathcal{V}_{Q}^{o2} = 1. \label{SQo}
\end{equation}

The ``quality" of the WWM is given in this model simply by
\begin{equation}
\mathcal{Q}_D = \mathcal{V}_D^o \;|\sin \Phi|. \label{QD}
\end{equation}
$\mathcal{Q}_D $ characterizes  ``how good'' the WWM is to establish
quantum correlations leading to the acquisition of WWI about the
quanton. Thus, the dependence of Eq. (\ref{QD}) on the entangling
phase can be understood. The dependence on $\mathcal{V}_D^o$ is also
important. As discussed in \cite{Jesus04}, as the detecton acquires
WWI about the quanton it also degrades its own fringe visibility.
This is due to the symmetric design of the SQDS, for which each
qubit is the WWM of the other. According to duality in this
reciprocal system, both qubits acquire WWI about each other and both
degrade their visibility. Thus, the initial visibilities for quanton
and detecton systems previous to their interaction act as limiting
factors of their subsequent mutual transfer of WWI.

Now, we compute $\Xi$ for the quanton, i.e.,
\begin{equation}
\Xi_Q^2 = \mathcal{P}_Q^2+\mathcal{Q}_D^2 (1-\mathcal{P}_Q^2).
\label{XiQ}
\end{equation}

The distinguishability $\mathcal{D}$ of the ways is given in Eq.
(50) of \cite{Jesus04} as
\begin{equation}
\mathcal{D}_Q = \text{Max} \left\{  \mathcal{P}_Q, \mathcal{R}_Q
\right\} , \label{SQDS10}
\end{equation}
where
\begin{equation}
\mathcal{R}_Q^2 = \mathcal{P}_Q^2 |\boldsymbol{s}_D^o|^2 +
\mathcal{Q}_D^2
 \left( 1-\mathcal{P}_Q^2  \right) .
\label{SQDS11rewritten}
\end{equation}
As can be seen from Eqs. (\ref{XiQ}) and (\ref{SQDS10}), $\Xi$
equals $\mathcal{D}$ in the six cases where $\mathcal{Q}$ or
$\mathcal{P}_Q$ or $|\boldsymbol{s}_D^{(0)}|$ reach their extreme
values 0 or 1. In between, the visibility limit $\mathcal{V}_\Xi$
 is always lesser than $\mathcal{V}_\mathcal{D}$. In order to show this, we define the deviation
$\Delta=\mathcal{V}_D^2-\mathcal{V}_{\Xi}^2$.  Inserting Eqs.
(\ref{SQDS10}) and (\ref{XiQ}) into Eqs. (\ref{defi1}) and
(\ref{defi2}), respectively, $\Delta$ can be written as
\begin{equation}
\begin{tabular}{ccc}
$\Delta =$ &\huge{\{}&
\begin{tabular}{ll}
$\mathcal{Q}^2 (1-\mathcal{P}_Q^2)$ \hspace{1cm} &$\mathcal{P}_Q
>
\mathcal{R}_Q$, \\
$\mathcal{P}_Q^2 \left( 1- |\boldsymbol{s}_D^o|^2\right)$
\hspace{1cm} &$\mathcal{P}_Q \le \mathcal{R}_Q$.
\end{tabular}
\end{tabular}
\label{tabla}
\end{equation}
Thus $\Delta\ge 0$ and $\mathcal{V}_{\Xi}\le
\mathcal{V}_\mathcal{D}$. In fact, as shown in the Appendix, Eqs.
(\ref{tabla}) coincide with Eq. (\ref{O49}).

\begin{figure}
\includegraphics{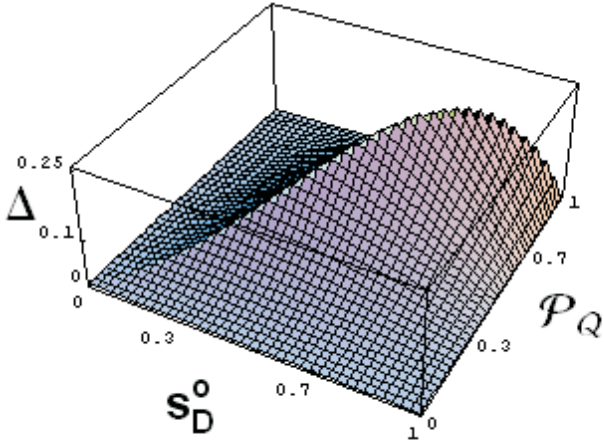}
\caption{\label{fig3}  Difference $\Delta$ between the squares of
the upper bounds of the visibility as a function of the norm of the
Bloch vector $|\boldsymbol{s}_D^{(0)}|$ and the quanton
predictability $\mathcal{P}_Q$. The plot is taken for a balanced
detecton $\mathcal{P}_D=0$ coupled maximally to the quanton
($\mathcal{Q}=|\boldsymbol{s}_D^{(0)}|$).}
\end{figure}

$\Delta$ is plotted in Fig. \ref{fig3} as a function of
$|\boldsymbol{s}_D^{(0)}|$ and $\mathcal{P}_Q$ for maximal coupling
($\sin \Phi=1$). We take for simplicity a balanced detecton
($\mathcal{P}_D=0$) so, according to Eqs. (\ref{SDo}) and
(\ref{QD}), $\mathcal{Q}=|\boldsymbol{s}_D^{(0)}|$.
As can be seen in the plot $\Delta=0$ for pure-state preparation
($|\boldsymbol{s}_D^{(0)}|=1$), since in this limit $\Xi$ always
equals $\mathcal{D}$. The opposite limit of a totally unpolarized
detecton ($|\boldsymbol{s}_D^{(0)}|=0$) leads to the limit of ``bad"
WWM ($\mathcal{Q}=0$) and $\Xi=\mathcal{D}$ since both quantities
equal $\mathcal{P}_Q$. Conversely, for $\mathcal{P}_\text{Q}=0$ we
have $\Xi=\mathcal{D}$ since both quantities equal $\mathcal{Q}$.
This reciprocity stems from the symmetric structure of the
right-hand side of Eq. (\ref{XiQ}) which yields $\mathcal{P}_Q$
($\mathcal{Q}$) when $\mathcal{Q}$ ($\mathcal{P}_Q$) vanish. Thus,
there is only deviation $\Delta \neq 0$ for unbalanced quanton
interferometers ($\mathcal{P}_Q\neq 0$) and mixed state preparation.
Finally, $\mathcal{P}_Q=1$ gives $\Xi=\mathcal{D}=1$ and the
interference fringes are totally degraded so $\Delta=0$. In between,
$\Delta$ can increase up to 0.25 for
$\mathcal{P}_Q=\mathcal{Q}=|\boldsymbol{s}_D^{(0)}|=0.7$.

We have proved the right hand side  of Eq. (\ref{3V}). Now we check
the left-hand side. The quanton visibility is given in Eq. (57) of
\cite{Jesus04} as
\begin{equation}
\mathcal{V}_Q = \mathcal{V}_Q^o \,\sqrt{\cos^2 \Phi +\mathcal{P}_D^2
\sin^2\Phi } . \label{SQDS13}
\end{equation}
Squaring and summing the above equation with Eq. (\ref{QD}) we have
\begin{equation}
\mathcal{Q}_D^2 +\frac{\mathcal{V}_Q^2}{\mathcal{V}_Q^{o2}} =
|\boldsymbol{s}_D^o|^2 \sin^2\Phi + \cos^2\Phi \leq 1 ,
\label{SQDS15}
\end{equation}
Now multiply Eq. (\ref{SQDS15}) by $(1-\mathcal{P}_Q^2)$ and
simplify using Eqs. (\ref{SQo}) and (\ref{XiQ}). We obtain
\begin{equation}
\mathcal{V}_Q^2 +\Xi_Q^2\le 1 , \label{O46}
\end{equation}
as we wanted to show.

\begin{figure}
\includegraphics[height=5cm]{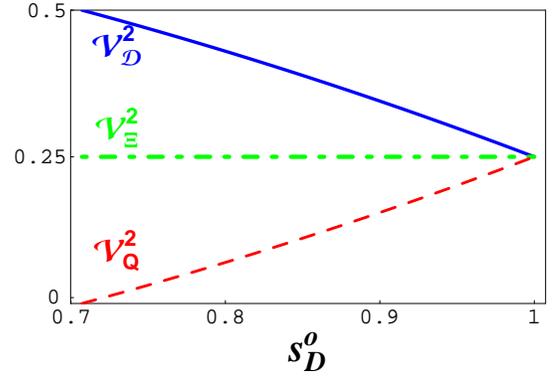}
\caption{\label{fig4}  $\mathcal{V}_\text{D}^2$ (solid line),
$\mathcal{V}_\Xi^2$ (dot-dashed line), and  $\mathcal{V}_\text{Q}^2$
(thin dashed line) as a function of the detecton's initial purity
$|\boldsymbol{s}_D^{(0)}|$. The plot is taken for
$\mathcal{V}_\text{D}^{o2}=\mathcal{P}_\text{Q}^2=0.5$ and a maximum
entangling phase $\Phi=\pi/2$.}
\end{figure}

This simple system illustrates different scenarios for Eq.
(\ref{3V}). First take $\mathcal{P}_D=0$. Then, according to Eqs.
(\ref{SDo}) and (\ref{QD}), $\mathcal{Q}_D$ is proportional to the
initial purity of the WWM given by the length of its Bloch vector,
while $\mathcal{V}_Q/\mathcal{V}_Q^o$ depends only on the entangling
phase. Lowering the initial purity of the WWM degrades the WWM's
ability to store WWI about the quanton. This action keeps
$\mathcal{V}_Q$ unchanged but lowers both $\mathcal{D}_Q$ and
$\Xi_Q$. Moreover, $\mathcal{D}_Q$ is lowered more than $\Xi_Q$, as
can be seen by comparison of Eqs. (\ref{XiQ}) and
(\ref{SQDS11rewritten}), which yields $\mathcal{V} \le
\mathcal{V}_{\Xi} \le \mathcal{V}_{\mathcal{D}}$. Now take
$\mathcal{V}_D^o$ constant and $\mathcal{P}_D\neq 0$, so degrading
the purity corresponds to lowering $\mathcal{P}_D$. In this case,
this action keeps $\mathcal{Q}_D$ unchanged, but degrades
$\mathcal{V}_Q/\mathcal{V}_Q^o$. Since $\mathcal{Q}_D$ and
$\mathcal{P}_Q$ stay constant $\Xi_Q$ remains invariant, as can be
seen from Eq. (\ref{XiQ}). However, $\mathcal{D}_Q$  degrades for
$\mathcal{R}_Q\ge \mathcal{P}_Q$ since its value depends in this
case explicitly on the value of the WWM's purity, as can be seen
from Eq. (\ref{SQDS11rewritten}). The condition $\mathcal{V} \le
\mathcal{V}_{\Xi} \le \mathcal{V}_{\mathcal{D}}$ is recovered, as
can be seen in Fig. \ref{fig4}. Again, a maximum value of
$\Delta=\mathcal{V}_{\mathcal{D}}^2-\mathcal{V}_{\Xi}^2=0.25$ can be
obtained as in the previous case of the balanced detecton situation.

\section{CONCLUSION}
\label{sectionConclusion}


$\mathcal{V}^2 +\Xi^2 \le 1$, like Eq. (\ref{O1}), is an inequality
quantifying duality. Like Eq. (\ref{O1}), it devolves the same
extreme cases given in Eqs. (\ref{boncho}) which constitutes the
usual statements about duality, i.e., concerning full WWI and no
fringe visibility (particlelike aspects), or perfect fringe
visibility and no WWI (wavelike aspects). In addition, both
inequalities also quantify intermediate situations (not specified by
the duality principle) where only partial WWI and partial fringe
visibility are possible. In these situations the two inequalities
yields two upper bounds for the visibility $\mathcal{V}_{\Xi}$ and
$\mathcal{V}_{\mathcal{D}}$. For balanced interferometers
($\mathcal{P}=0$) $\mathcal{V}_{\Xi}=\mathcal{V}_{\mathcal{D}}$,
since in this case $\Xi=\mathcal{Q}=\mathcal{D}$. For pure states,
we also find that both visibility bounds coincide. However, for
mixed states and unbalanced interferometers we obtain a remarkable
result. We find that the visibility bound
 $\mathcal{V}_{\Xi}$ can be lower than $\mathcal{V}_{\mathcal{D}}$.
Thus, $\mathcal{V}^2 +\Xi^2 \le 1$ can be more stringent than
$\mathcal{V}^2 +\mathcal{D}^2 \le 1$.

Note that $\mathcal{D}$ is the largest possible value of the
different knowledge $\mathcal{K}_W$ associated with a concrete
betting strategy \cite{Englert96,Englert99}. Thus,
$\mathcal{V}_{\mathcal{D}}$ in Eq. (\ref{VVD}) is just the lowest
upper bound to the visibility from all the bounds
$\sqrt{1-\mathcal{K}_W^2}$ associated with different measurement
choices of WWM's observables $W$. On the other hand, since
$\mathcal{V}\leq \mathcal{V}_{\mathcal{D}}$ does not saturate for
mixed-state preparation, lower minima such as $\mathcal{V}_{\Xi}$
can still be found. The bound $\mathcal{V}_{\Xi}$ is not associated
in general with the betting strategy mentioned before. Nevertheless,
like $\mathcal{V}_{\mathcal{D}}$, $\mathcal{V}_{\Xi}$ is  a
manifestation of the quantum correlations established between the
quanton and the WWM, derived through the algebraic properties of
their Hilbert space.

We have particularized the results to a concrete example: the
symmetric quanton-detecton system, which illustrates the different
quantities introduced in the formalism. The inequalities
$\mathcal{V} \le \mathcal{V}_{\Xi} \le \mathcal{V}_{\mathcal{D}}$
are shown to hold for this particular system. Actually, the
difference between the visibility limits $\mathcal{V}_{\Xi}^2$ and
$\mathcal{V}_{\mathcal{D}}^2$ in this system is significant and can
reach a value of 0.25.

%

\begin{acknowledgments}
This research was supported by a Return Program from the
Consejer\'{i}a de Educaci\'on y Ciencia de la Junta de Andaluc\'{i}a
in Spain.
\end{acknowledgments}
%

\appendix*
\section{}
Here we show  the connection between Eqs. (\ref{tabla}) and the more
general expression given in Eq. (\ref{O49}). In order to do that,
consider the mixed state
\begin{equation}
\rho_D^0=D_1 | \uparrow \rangle \langle \uparrow | + D_2 |
\downarrow \rangle \langle \downarrow |,
\end{equation}
where
\begin{equation}
D_{1,2}= \frac{1}{2} (1\pm S_{Dz}^0), \label{D12}
\end{equation}
and $\{ | \uparrow \rangle, | \downarrow \rangle \}$ are the usual
eigenvectors of $\sigma_{Dz}$. The quantity
$\chi=\mathcal{D}^2/\Xi^2$ can be written in terms of $\Delta$ as
\begin{equation}
\chi=1-\frac{\Delta}{\Xi^2}.
 \label{chiDelta}
\end{equation}
With the help of the three previous equations, Eqs. (\ref{tabla})
can be written
\begin{equation}
\begin{tabular}{ccc}
$\chi =$&\Huge{\{}&
\begin{tabular}{ll}
$\mathcal{P}_Q^2/\Xi^2$ \hspace{1cm} &$\mathcal{P}_Q >
\mathcal{R}_Q$, \\
\\
$1-(4D_1D_2 \mathcal{P}_Q^2/ \Xi^2) $ \hspace{1cm} &$\mathcal{P}_Q
\le \mathcal{R}_Q$.
\end{tabular}
\end{tabular}
\label{tablachi}
\end{equation}
which coincides with Eq. (\ref{O49}).

%


%
%
%


\begin{references}
\bibitem{comentario inversion}
As noted in \cite{Englert96b}, there is no need to consider a more
generic Bloch vector than that given in Eq. (\ref{2}), since this
just amounts to a redefinition of the operators $V_{\pm\pm}$ in
(\ref{O6}).
\bibitem{comentario D=1}
We exclude the trivial case $\mathcal{D}=1$, where the three
visibilities of Eq. (\ref{3V}) vanish.
\end{references}
\end{document}